# Temperature and high-pressure dependent X-ray absorption of $SmNiO_3$ at the Ni K- and Sm $L_3$- edges

Néstor E. Massa*,[1] Aline Y. Ramos,[2] Helio C. N. Tolentino,[2] Narcizo M. Sousa-Neto,[3] Jairo Fonseca Jr.,[3] and José Antonio Alonso.[4]

[1]Laboratorio Nacional de Investigación y Servicios en Espectroscopia Óptica-Centro CEQUINOR, Universidad Nacional de La Plata, C.C. 962, 1900 La Plata, Argentina,

[2]Institut Néel, CNRS et Université Joseph Fourier, BP 166, F-38042 Grenoble Cedex 9, France,

[3]Laboratório Nacional de Luz Síncrotron - P.O. Box 6192, 13084-971, Campinas, São Paulo, Brazil,

[4]Instituto de Ciencia de Materiales de Madrid, CSIC, Cantoblanco, E-28049 Madrid, Spain.

•e-mail: neemmassa@gmail.com

# ABSTRACT

We report on XANES and EXAFS measurements of $SmNiO_3$ from 20 K to 600 K and up to 38 GPa at the Ni K- and Sm $L_3$- edges. A multiple component pre- Ni K-edge tail is understood originating from 1s transitions to 3d-4p states while a post-edge shoulder increases distinctively smooth, at about the insulator to metal phase transition ($T_{IM}$), due to the reduction of electron-phonon interactions as the Ni 3d and O 2p band overlap triggers the metallic phase. This effect is concomitant with pressure induced Ni-O-Ni angle increments toward more symmetric $Ni^{3+}$ octahedra of the rhombohedral $R\bar{3}c$ space group. Room temperature pressure dependent Ni white line peak energies have an abrupt ~3.10±0.04 GPa Pa valence discontinuity from non-equivalent $Ni^{3+\delta}$ + $Ni^{3-\delta}$ charge disproportionate net unresolved absorber turning at ~$T_{IM}$ into $Ni^{3+}$ of the orthorhombic P*bnm* metal oxide phase. At 20 K the overall white line response, still distinctive at $T_{IM}$ ~8.1±0.6 GPa is much smoother due to localization. Octahedral bond contraction up to 38 GPa and at 300 K and 20 K show breaks in its monotonic increase at the different structural changes. The Sm $L_3$-edge does not show distinctive behaviors either at 300 K or 20 K up about 35 GPa but the perovskite Sm cage, coordinated to eight oxygen atoms, undergoes strong uneven bond contractions at intermediate pressures where we found coexistence of octahedral and rhombohedral superexchange angle distortions. We found that the white line pressure dependent anomaly may be used as an accurate alternative for delineating pressure-temperature phase diagrams.

**Pacs number(s):**

X-ray absorption spectra--78.70.Dm

High-pressure effects in solids and liquids--62.50.-p; Crystallographic aspects of phase transformations--61.50.Ks (Pressure treatment--81.40.Vw)

Metal-insulator transitions and other electronic transitions--71.30.+h;

Strongly correlated electron systems--71.27.+a

# INTRODUCTION

$RNiO_3$ (R=Rare Earth) is a distinctive family of compounds in which a sharp temperature-driven insulator to metal phase transition is found at $T_{IM}$.[1] Except $LaNiO_3$, that remains metallic at all temperatures in a rhombohedral distortion, the series obtained by Rare Earth replacement has an insulating phase in which the Ni $e_g$- O 2p hybridization decreases as $T_{IM}$ increases. Above $T_{IM}$ they crystallize in the $GdFeO_3$-type orthorhombic paramagnetic distorted perovskite structure P*bnm*.[2]. Assignment of the orthorhombic P*bnm* space group by long-range X-ray diffraction measurements were concomitant with a weak reduction of the unit cell volume when passing into the metal phase. The undergoing process is a first order phase transition consequence of slight increasing the Ni-O bond length and a simultaneous decrease of the Ni-O-Ni angle. The latter being the driving force of the transition as it governs the transfer integral between Ni $e_g$ and O 2p orbitals. In $RNiO_3$ Ni is in a low spin electronic state.[3,4,5]

Neutron and synchrotron diffraction data from $YNiO_3$ provided the first evidence of changes in the crystal symmetry at the metal-insulator transition from high temperature orthorhombic metallic to monoclinic insulating due to charge disproportionation ($2Ni^{3+} \rightarrow Ni^{3+\delta}+Ni^{3-\delta}$) at the opening of the gap. These correspond to two alternating $NiO_6$ octahedra associated with an expanded (Ni1) and a contracted (Ni2) Ni-O bond and a magnetic structure [k=(1/2,0,1/2)] with unequal moments at independent Ni1 and Ni2 sites.[6]

More subtle local structure measurements showed this extended picture as a $P2_1/n$ monoclinic distortion in the insulating phase of $NdNiO_3$ (Ref. 7). It might also be though as a Jahn-Teller like distortion at one of the two non-equivalent Ni sites creating charge fluctuations in the metallic phase. Conversely, since $Ni^{2+}$-like ions are not Jahn-Teller active, the expected small polarons appear as consequence of the suppression of the $Ni^{3+}$-like Jahn-Teller distortion and conduction electrons will then move through the lattice destroying the quasi-Jahn-Teller distortion at the Ni sites.[8, 9]

Diffraction measurements from single crystals also showed the presence of the monoclinic P21/n phase in $PrNiO_3$ (Ref.10) alongside X-ray absorption near edge (XANES) measurements requiring the monoclinic distortion to explain the measured spectra.[11] Nonetheless, the overall medium and long-range structure is quite well reproduced considering only the orthorhombic symmetry for a single Ni site. The splitting takes place at the local level turning on a weak monoclinic distortion that materializes in the short-range scale.[6]

The complete physical scenario for this family of compounds emerged from fluorescence yield X-ray absorption confirming that at the insulating lower temperature charge-disproportionate state there is a rare-earth-dependent splitting of the Ni- K edge. The fluorescent band splitting is five times larger for $LuNiO_3$ [2.5(1) eV] than for $PrNiO_3$ [0.5(3) eV] pointing to decreasing charge transfer between $Ni^{3+\delta}$ and $Ni^{3-\delta}$ as the itinerant limit is neared. It also supports the persistence of dynamic $Ni^{3+\delta}/Ni^{3-\delta}$ charge fluctuations in the metallic phase.[12]

Unusually large $O^{16}$–$O^{18}$ isotope shifts in the metal-insulator transition temperature $T_{IM}$ indicates that, indeed, the mechanism of the transition in $RNiO_3$ involves strong electron-lattice interactions.[13, 14] Hopping conduction electrons move through the lattice in a similar way as $e_g$ small polarons in Mn hole doped perovskites.[15] The agreement found between the small polaron theory and the experimental optical conductivity of $RNiO_3$ (R=Rare Earth≠ La) confirms electron-phonon interactions as a determinant factor of phonons participating in the transport properties of these compounds. According to this, the metal-insulator transition at $T_{IM}$ is directly related to self-trapped electrons in a polaronic medium.[16. 17] At about 100 meV small charge transfer gap appears below $T_{IM}$.[18]

In the larger Rare Earth, (Nd or Pr) the sharp change in conductivity at $T_{IM}$ coincides with a Ni magnetic moment net antiferromagnetic ordering. Smaller size rare earth induces distortions unfolding an ion size dependent intermediate paramagnetic phase between the high temperature metallic and the truly magnetic ordered. In those last compounds, the temperature of the metal-insulator transition is different from the Néel temperature so that the magnetic ordering now develops upon cooling in an insulating matrix.[20] The temperature of $T_{IM}$ is higher decreasing the Rare Earth ion size due to an increment in the topological deviates from the ideal perovskite structure.[20] The same non-collinear magnetic structure is however shared by all $RNiO_3$ (R= Rare Earth≠ La)). The magnetic ground state consists of an unusual antiferromagnetic structure, with **k**=(1/2,0,1/2) relative to the orthorhombic crystal cell.[4, 21]

It is against this background that earlier pressure dependent transport measurements were carried out by Obradors et al[22] and Canfield et al[23] Both groups coincide in reporting a negative slope for $-dT_{IM}/dp$ in $RNiO_3$ (R=Rare Earth≠ La). Medarde et al studied by neutron-diffraction the high pressure metallization process in $PrNiO_3$ under moderate pressures[24] and, more recently, Lengsdorf et al reported crystallographic work in $EuNiO_3$ under hydrostatic pressure.[25] The sequence from orthorhombic to rhombohedral symmetry has also been observed by angle-resolved x-ray diffraction measurements at 11 GPa and 6 GPa in $NdNiO_3$ and $PrNiO_3$, respectively.[26]

In order to propose a general model, we look at $SmNiO_3$ which is an intermediate nickelate sharing all the above properties.[2] $SmNiO_3$ undergoes a metal-insulator phase transition at ambient pressure at $T_{IM}$=400K (Ref. 27) and the Néel insulating phase stabilizes below $T_N$=225K (Ref. 15) .

The primary consequence of applying quasihydrostatic pressure to a compound, as in our case a perovskite, is the reduction of interatomic distances, the progressive bond shortening, and the modification of superexchange M-O-M angles. Continuous pressure change as thermodynamic external dynamical variable is a unique tool for probing the relationship between structural sublattice distortions and electronic properties. It triggers increments in band hybridization and the eventual emerging of a metallic state by valence and conduction band superposition.

Structural changes in the lattice parameter have been observed in $SmNiO_3$ under pressure at room temperature by means of angle-resolved X-ray difffraction. At around 2.5 GPa the unit-cell parameters are reported to undergo a small abrupt contraction that it is the onset of the pressure-induced metallization earlier found by infrared spectroscopy at ~1.9 GPa.[29] This is a consequence of the increase of the superexchange angle Ni-O-Ni from 153.4°at ambient pressure to more than 156° at 3 GPa. A further increase of the applied pressure triggers a region of orthorhombic-rombohedral phase coexistence that then leads to the less distorted metallic rhombohedral ($R\bar{3}c$) phase shared with $LaNiO_3$ (Ref. 29). Pressure induced increments in the Ni-O-Ni angle with simultaneous diminution of the Ni-O distances, given the negative slope $-dT_{MI}/dp$ in $RNiO_3$ (R=Rare Earth≠ La), eventually closes the insulating gap.[22,23] As it was also found recently in $YNiO_3$ , these changes represent the interplay of the band width W, the charge energy transfer Δ, and the Coulomb repulsion energy

U by which an increment in W and thus W/U, (the onsite U Coulomb energy is not expected to change significantly because it depends on more localized 3d orbitals) yields a reduction of the temperature at which insulator metal transition takes place.[32]

Here we report on temperature and pressure induced dependences in X-ray absorption (XAS) at the Ni K- and Sm $L_3$- edge of $SmNiO_3$. The Ni K-edge results from 20 K to 600 K at ambient pressure are used to develop models for the analysis of the pressure dependent counterpart. Pressure dependence X-ray absorption of $SmNiO_3$ up to 38 GPa and at 300 K and 20 K was used to explore XANES (X-ray absorption near edge structure) profiles below and above the metal-insulator and the lower temperature antiferromagnetic phase transition. We found that upon applying pressure the change of the Ni K-edge white line peak energies are a good indicator of the $Ni^{3+}$ metallic phase onset suggesting a handy tool to map pressure-temperature phase diagrams in disproportionated compounds. At 20 K, high-pressure dependent measurements show a rather smooth evolution of the Ni white line peak energy suggesting dynamical carrier localization in the paramagnetic phase before the onset of the insulator to metal transition at ~8.1±0.6 GPa.

The octahedral contraction, mirroring the increment in energy of the white line has been also quantitatively studied at 300 K and 20 K from fits to the pseudo distribution function of EXAFS data. The pressure induced relative changes in the Ni-O bonds yields slope changes whose onset is related to pressures at which the system undergoes structural modifications. These take place at $T_{IM}$ (~3.10±0.04 GPa at 300 K and ~8.1±0.6 GPa at 20 K); at the intermediate phase where the superexchange angle shares both orthorhombic and rhombohedral incremental distortions, and in the mostly rhombohedral environment at the highest pressures reached in our experiments. Overall, our measurements bring up one more time the role of octahedral tilting and p-d hybridization in correlated oxides.

We also performed detailed temperature and pressure dependent measurements at the Sm $L_3$-edge. We found that these spectra do not have distinctive band profile changes, either at 300 K or 20 K, and up about 35 GPa, confirming the absence of a major role for the Rare Earth beyond the lattice distortion introduced by ion size underlying the metal to insulator mechanism. On the other hand, our probe gives evidence of strong distortions in the Sm cage at the same pressures where we detected mixed superexchange angles sharing either orthorhombic or rhombohedral topologies.[26]

## SAMPLE PREPARATION AND EXPERIMENTAL DETAILS

Powder $SmNiO_3$ was prepared by a liquid-mixture technique. A stoichiometric mixture of metal nitrates was solved in citric acid and the citrate solution was then slowly decomposed at temperatures up to 600 °C. This black precursor was heated at 1000 °C under a 200 bars oxygen atmosphere for 12 h and cooled at room temperature, giving a polycrystalline powder. The sample was then characterized by x-ray and neutron powder diffraction, thermogravimetric analysis, differential scanning calorimetry, and resistivity measurements.[33]

Transmission-mode x-ray absorption measurements were done at the D04B-XAFS1 beamline in the Brazilian Laboratory for Synchrotron Light (Laboratorio Nacional de Luz Sincrotron-LNLS).[32] Channel-cut Si monochromators 220 and 111 reflections were utilized around Ni *K*-edge (~8345 eV) and Sm $L_3$-edge (~6716.2 eV). Low temperature measurements in the 20 K to 300 K interval were made using $SmNiO_3$ films deposited on a silicon nitride $Si_3N_4$ membrane mounted in the cold finger of a closed cycle He refrigerator. For the high temperature runs, between 300 K and 600 K, samples were prepared by mixing powder of $SmNiO_3$ with boron nitride and pressing the mixture into a pellet. Different pellets were prepared with amounts of $SmNiO_3$ and BN optimizing absorption edges jump. The samples were placed inside a home-made furnace where the temperature was varied.

High-pressure measurements in a quasi-hydrostatic environment were done at the dispersive beamline D06A-DXAS of the LNLS. Its Si(111) bent monochromator allows very high energy stability focusing into a 150 µ Gaussian spot at the sample position.[33] The reference flux, $I_0$, was measured with a flat piece of plastic in order to simulate the average attenuation. X-ray absorption were measured at 300 K and 20 K accumulating runs in the 100-200 range with the cell mounted on the cold finger of a He closed cycle refrigerator.

We have also reproduced some of our results at room temperature with the pressure cell mounted on a goniometer because close Ni-O bond lengths differences appear as static disorder due to the limited EXAFS *k* range reducing the resolution in the R space Fourier analysis.[34]

All measurements have been done in the transmission mode. We calibrated our spectra of $SmNiO_3$ using as standards a metal Fe sheet for Sm and a metal Ni sheet for Ni. This means that for the spectrometer resolution in our experimental configuration the precision of the pressure calibration was ~0.2 GPa.

The pressure cell was loaded with silicone oil embedded with $SmNiO_3$ 5µ grain powder using 250µm thick stainless steel gaskets pre-indented to approximately 60µm thick. That mixture was used for filling the gasket hole located between the two diamond anvils of a He gas driven membrane cell. To reduce the diamond absorption, thus also allowing the measurement of the Rare Earth $L_3$ edge, we used perforated anvils with 350μm cullet size. Pressure was calibrated in situ from temperature corrected band frequency shifts of the ruby laser excited R1 and R2 fluorescence lines.[35, 36, 37]

We found that at room temperature the use of silicone oil as transmitting medium yielded a good quasi-hydrostatic environment up to about 10 GPa while at higher pressures the fluorescence lines bandwidth increased echoing non hydrostatic departures.[38] Error bars where then estimated ±0.5 GPa for those points below 10 GPa while at higher pressures, it increased up to about ± 1.0 GPa consequent with the extra broadening. Nonetheless, as may be seen below, since XAS probes an effective average contribution from all micro-grains, this makes our set up not too dependent of the actual environment created by the pressure transmitting medium; this makes consistent measurements up to ~38 GPa.[30] possible.
All XAS data were analyzed using the Atoms-Athena-Artemis software package.[39]

## RESULTS AND DISCUSSION

### *i) Temperature dependence of the Ni K-edge in $SmNiO_3$ at ambient pressure.*

Figure 1 shows the temperature dependence of the Ni K-edge XAS spectra from 20 K to 600 K result of three scans recorded and averaged to improve the signal to noise ratio at every shown temperature. They were normalized using standard procedures.[39] As it has been reported for other members of the $RNiO_3$ (R≠ La) family[11] there are three main XANES features to follow as the material goes through the lower temperature antiferromagnetic ordering at $T_{Néel,}$, to the intermediate paramagnetic state, and into the insulator- metal transition at $T_{IM}$ ~400 K. They are usually named *A, B, C; A* stands for any pre-edge tail-like feature that in our case, Fig. 2 (A), has multiple components.

We adopted a Weibull profile for fitting of the white line *B* because its slight asymmetry is better reproduced by the sigmoidal extra "c" fitting parameter improving the deconvolution of pre-edge features. Similar earlier less resolved weaker features were assigned in $PrNiO_3$ to mainly dipolar transitions between metal 1s states and hybridized metal 3d-O2p bands [11, 40]. In our case, the slope change in the while line shape at ~8344 eV (Fig. 2 (A)) suggests two components, $A_1$ and $A_2$ that fitted with two Lorenztians, are reminiscent to the two bands discussed for NiO in the same energy range. Within this framework, the pre-edge $A_1$ (Fig. 2 (A), inset (a)) is assigned to Ni transitions from 1s state to empty 3d ($\Delta$l= ±1) likely mixed with 4p levels result of p-d hybrization with nearest neighbor oxygen.[41] Pre-edge $A_2$ is understood coming from non-local dipole-like excitations due to transitions to d sates of neighboring Ni ions promoted by hybridization with the transition metal absorbing p states. That is, they involve higher coordination shells in the Ni-O-Ni interactions.[40]

We find, however, what might be interpreted as a quite the opposite behavior. Beyond a faint temperature dependence suggested in Fig. 2(A) inset (b), pre-edges do not seem to experience profile changes either by charge disproportion in the two Ni nonequivalent sites[6] nor by freer carriers above 400 K. It is known that strong hybridization among the Ni3d:eg and O2p:s bands leads to the partial screening distortion by holes transferred from the 3d to the 2p orbital , and remarkable, these are the same end states claimed as origin of the pre-edge structure. The octahedron off-center displacement and stretching decreases the local symmetry (non-centrosymmetric) allowing the degree of p-d mixing raising the pre-edge XANES peaks.[34,42,43]

We then conclude that the origin of those features might be in a localized electron phase co-existing in the metallic matrix above $T_{IM}$ where the conduction electrons role would be preventing the charge-order state. That is, split sites remaining fluctuating associated with local distortion of $NiO_6$ octahedra.[34, 44]

In sharp contrast with the lack of response to the onset of the metallic phase by pre-edge features, post-edge shoulder *C* (Fig.2 (B)) at ~ 8358 eV grows up increasing temperature. It originates from multiple scattering contributions by local octahedral $NiO_6$ changes in the Rare Earth perovskite cage representing freer carriers under reduced electron-phonon interactions. The onset of the passage to metallic phase in $SmNiO_3$ may associated with far infrared reflectivity where vibrational modes become distinctively screened as a consequence of a non-negligible reduction in the coupling of electronic degrees of freedom to local lattice distortions.[16,17]

A shoulder *C* sharp change passing from insulator to metal regime has been already observed in $PrNiO_3$ (Ref.11) and in $NdNiO_3$ (Ref.45). These two compounds lack of the intermediate paramagnetic phase. Shoulder *C* in $SmNiO_3$ (Fig. 2 (B)) does not change as abruptly when going from the insulating monoclinic distorted phase to the conducting orthorhombic: Rather, the absorption gradually increases starting about 300 K in the intermediate paramagnetic phase (Fig. 2 (B), inset) in presence of $Ni^{3+\delta}$ and expanded $Ni^{3-\delta}$ sites recalling $Ni^{3+\delta}/Ni^{3-\delta}$ charge fluctuations[31] as a source of gradual change incrementing the conductivity.

We then conclude that the change in the absorption cross section at the post-edge responds to contributions from charge disproportion dynamics and local environment inhomogeneities at both sides of $T_{IM}$ that can be used as a quantitative determination of the passage to the metal oxide phase.[30] In turn, the origin of the intermediate paramagnetic phase might be though as being induced by inhomogeneous lattice distortions due to larger charge disproportions. It is known that for $PrNiO_3$, $N_1$ and $N_2$ valences are very close to trivalent $Ni^{3+}$ while $SmNiO_3$ is already in an intermediate stage with finite $\delta$ closer to a tetravalent/divalent $Ni^{4+}$-$Ni^{2+}$ valence picture reported in smaller Rare Earths as for $LuNiO_3$.(Ref.12) This sets apart nickelates in which $T_{IM}$ coincides with $T_{Néel}$ ($PrNiO_3$, $NdNiO_3$) from those ($RNiO_3$, R$\neq$ La, Pr, Nd) that have $T_{IM} > T_{Néel}$ at well differentiated temperatures.

Our temperature dependent EXAFS (extended X-ray absorption fine structure) spectra (Fig. 1) have similar features as earlier runs for other Rare Earth nickelates.[34] In particular, the Fourier transform amplitude in the magnetic ordered, paramagnetic, and metallic phases shows a distinctive asymmetry in the band profile of the pseudoradial distribution function of the Ni first neighbors oxygen coordination shell.[34] This asymmetric deviation tend to disappear in the metallic phase suggesting a reduction in the Ni splitting. i.e.; turning into a unique single $Ni^{3+}$ environment signaling the passage from P21/n space group to orthorhombic P*bnm*. Then, a first shell may be modeled introducing explicitly an apical (two equal O-Ni nearest neighbor distances) and a basal plane (four equal Ni-O bonds) and then fitted to the experimental temperature dependent radial pseudo radial distribution. We found that the calculated temperature dependent of bond lengths (Fig. 1, inset) are in very good agreement with earlier crystallographic data. They have an smooth behavior in the insulating phase increasing at higher temperatures in the metallic phase.[19

,46] These bonds were then used in the following sections as atmospheric pressure reference in our analysis of the high pressure data.

It is also worth mentioning that we also found a slight change at about $T_N$ ~225 K showing, in agreement with transport measurements by Zhou et al.[46] a weak lattice interaction (spin-phonon) at the temperature where the Ni moments align.

Nonetheless as was already commented above a limiting factor when performing high pressure absorption measurements is the reduction of the measuring *k*-range consequence of the pressure cell mounted at the beamline focal point. The measured spectra are comparable to long-range X-ray diffractograms yielding "average" results for octahedral bond lengths as if the lattice were labelled by orthorhombic P*bnm* ignoring departures from the ideal $Ni^{3+}$ valence.

### ii) Temperature dependence of the Sm $L_3$ –edge in $SmNiO_3$ at ambient pressure.

In Fig. 3 we show our temperature dependent XAS spectra at the $L_3$ Sm edge of $SmNiO_3$. The sharpness of the white line, typical of Sm(III), corresponds to the 2p → 5d transition in $Sm^{3+}$.The temperature dependence does not show abrupt changes that may be directly associated with phase transitions in $SmNiO_3$ but rather the complete range of our study is dominated by the macroscopic lattice distortion triggered by Sm reducing the ideal cubic perovskite topology. The lattice structure becomes unstable due to the ion size coercing distorted $NiO_6$ octahedra into tilt and rotation to fit the otherwise empty cubic space. We do however note, that in that process there is also a small change in the relative intensity of the white line (Fig. 3, inset a) suggesting that the number of unoccupied states in a 5d-band are reduced in passing to the orthorhombic metallic phase.

Regardless, our main purpose in doing these measurements is to provide a more quantitative view of the dodecahedral Sm first nearest neighbors scattering paths for building a realistic model to analyze the absorption spectra under pressure. In this order after an EFFF9 analysis[39] we found five Sm-O scattering paths between 1 Å and 3 Å globally considered as subshells of a first shell in the R space of the Sm pseudoradial distribution function yielding a set of five bonds (Fig. 3, inset b) in agreement with the published crystallographic data.[47]

iii) ***High pressure dependence of the Ni K-edge in $SmNiO_3$ at 300 K***

Fig. 4 (A) shows the X-ray absorption XAS spectra at the Ni K-edge in $SmNiO_3$ at 300 K as function of pressure. Every feature is altered even under the lowest applied pressures. A remarkable behavior is observed at the post-edge *C* shoulder (Fig. 4 (B)) where one is able to distinguish the 300 K pressure induced passage from the insulator to the metal state starting in the intermediate paramagnetic phase into a phase distortion in which carriers are freer. Increasing the pressure further, the absorption increases smoothly reaching metallic rhombohedral symmetry in which carriers experience weaker electron-phonon interactions in a metal oxide environment (Fig. 4 (B), inset).

Pre-edge *A* bands at 300 K, Fig, 5, seem to develop, within their weakness, sharper profiles upon applying pressure relative increasing their intensity and half width at half maximum.[49]

On the other hand, the evolution in the lattice distortion may be traced by following the energy peaking of the white line. It is known that absorption edges are highly sensitive to the absorber oxidation state. It is also known that one may expect an increase in the peak energy due to octahedral contraction. Higher oxidation-states require more energetic X-ray to excite the core electron due to less-shielded nucleus carrying a higher effective charge.[30] It implies shorter bond lengths prompting edge energies increments.

Then, exploiting both facts, we may understand the abrupt change in white line peak position, shown in Fig. 6 (A), as the pressure driven passage from two unresolved $Ni^{3+\delta}$ and $Ni^{3-\delta}$, and non-equivalent absorbers, at ~3.10±0.04 GPa to an effective $Ni^{3+}$ oxidation state of the metallic phase. There is a small change in the unit cell volume due to the Ni-O bond decrease induced by electronic delocalization at $T_{IM}$ (Ref. 15).

Fourier transform the EXAFS spectra, still at ambient temperature, gives another quantitative view. This is shown in Fig. 6 (B) as pseudoradial distribution functions (no phase shift correction) of the pressure dependence of Ni-O first shell. As it was already anticipated dealing with the temperature dependent case, in the high-pressure counterpart the first Ni shell (Fig. 6 (B)) involving Ni-O oxygen bonds does not show structure due to the limitation in *k* space. The restricted EXAFS *k*-range characterizes a low-resolution study in R-space. The Ni-O bond lengths are not distinguishable and bond length differences within the two Ni sites appears as a static disorder contribution to the total bond length dispersion. We then fitted the pseudo radial distribution function with the model found in the temperature dependent counterpart considering apical (two nearest neighbors) and basal (four equal Ni-O bonds), now, with correlated distances. Intuitively, short bonds should be less reduced than longer ones by the application of an external pressure.[49] Accordingly, the parameters changing in the fits common to both sets are the edge energy shift $E_o$, bond length R, and the Debye-Waller factor $\sigma^{**2}$.

The octahedral bond contraction, Fig. 6 (C), shows structural modification at slope changes breaking its monotonic pressure dependence.

This allows to link the anomaly at $T_{IM}$, Fig 6(A) to structural changes across the metal-insulator transition. The pseudo radial distribution analysis, Fig 6(C) shows the abrupt change at $T_{IM}$ in Fig. 6(A) as a steep drop reflecting a rather strong bond contraction below 3 GPa. The inhomogeneous charge distribution at $T_{IM}$ was identified in $RNiO_3$ (R= Ho, Y, Er, Lu) by the symmetry breaking of the oxygen positions implying simultaneous strong octahedral distortions.[50] This crystallographic measurements showed that the charge disproportion completely vanishes at the insulator to metal transition with an abrupt convergence of the two disproportionated sets,Ni1 and Ni2 [5, 12]. The three-bond lengths corresponding to the monoclinic distortion in the insulating phase changing to a unique three Ni-O distances of the orthorhombic metallic [50].

Further increasing pressure induces a region of coexistence of orthorhombic to rhombohedral distortion that coincident with diffraction measurements[26] suggests a more or less deformed lattice evolving toward a more symmetric topology through variations and coexistence of the superexchange angle.

Higher pressure triggers another break at above ~30 GPa indicative of entering a mostly rhombohedral R-3c region. The angular lattice crossover goes from the 152° orthorhombic to the 165° rhombohedral onset of the lattice symmetry shared with atmospheric pressure metallic $LaNiO_3$ (Ref 29).

iv) ***High pressure dependence of the Ni K-edge in $SmNiO_3$ at 20 K***

We recognize at 20 K, Fig. 7(A), the same sequence of pressure induced changes in the Ni-O bond contraction. Several indicators suggest, however, a more static and localized condition for carriers.

Fig. 5 shows pre-edge features having longer energy span at 20 K. Although noisier, for a better fit it is necessary to consider a possible $A_1$ split into two lorenztians assigned to $e_g$ and $t_{2g}$ states. The second band, $A_{2,}$ remains within the relative increased parameters found for the 300 K runs. The post-edge *C*, Fig. 7 (B), on the other hand, clearly denotes the onset and gradual change toward the more conducting environment with traces marking the passage through the already known intermediate steps of the different structural rearrangements in metallic phases.

The most remarkable effect in this series at low temperature is, however, due to localization and the environment created by the $Ni^{3+\delta}$ and $Ni^{3-\delta}$ monoclinic distortion in conjunction with the unresolved Ni edge. The temperature dependent localization that, on cooling at atmospheric pressure, drives the insulator to metal transition[16] is enhanced by the application of external pressure. This increases carrier correlation and thus Coulomb interaction that by hybridization governs interatomic distances as the volume decreases. We know that in $RNiO_3$ (R=Rare Earth≠La) the small change in volume at $T_{IM}$ implies a first order transition in which metallization takes places due to the electron correlation and structural distortion. [2] This results in a kind of end product reordering of the bands according to the U/W<1 ratio (W being the factor susceptible to pressure and temperature changes).

At 20 K, the white line peak energies are found to show a much smooth evolution toward the point in which unresolved absorbers $Ni^{3+\delta}$ and $Ni^{3-\delta}$ change its weigh in favor of the $N^{3+}$ oxidation state. We individualize this point as a change in the slope, onset a white line peak energy small increase toward higher symmetry $Ni^{3+}$ octahedra (Fig. 8 (A)). The slope change in Fig. 8(A) and Fig.8(C) signals the passage to the conducting phase at ~8.1±0.6 GPa.

That is, these measurements yield a low temperature scenario in which a pressure driven disproportionated insulating state undergoes a more continuous W pressure induced band broadening. This then may be though as if the first order character of the $T_{IM}$ transition turn into a mix behavior adding a lattice displacive constituent commonly found in almost all order-disorder phase transitions in innumerable simple distorted perovskite oxides. The mixed character causes a more continuous second-order-like evolution of the distortion of Ni-O bonds. Pressure induced structural changes are concomitant with electronic localization mimicking the known rare ion size induced lattice distortion. The emerging of the conducting phase with higher symmetry $Ni^{3+}$ cations in a single octahedral scenario above 30 GPa would be associated to this The gradually straightening of the $NiO_6$ octahedra governs the 3d-2p overlap.

Then, our new three points in the phase diagram, Fig. 10, are in essence consequence of the Mott-like transition. [51] Localizing carriers lowers the energy of the system opening a gap. This can be upset by applying pressure broadening the states, reducing ion bond distance, and thus increasing hybridization. In $RNiO_3$ (R=Rare Earth≠La), the nature of $T_{IM}$ transition is the result of the combined effect of the Mott band approach and bond disproportion associated with disproportionated sites.[52]

The 20 K behavior of the low temperature white line peak and the stronger anomaly at 300 K may then be seen as an alternative for mapping the pressure phase diagram of $SmNiO_3$. Consequent to an unresolved absorber and octahedral contraction it converges into P–T points identified as those at which the insulator to metal transition phase transition takes place.

It is also worth noting that our ~8.1±0.6 GPa for $T_{IM}$ at 20 K suggests a phase diagram for $SmNiO_3$ closer and common to the known from resistance measurements earlier reported by Obrador et al[22] and Canfield et al[23] for $NdNiO_3$ and $PrNiO_3$ . It suggests a behavior shared by all members of this family of compounds. Our results ought to be viewed as a refinement of 6.2 GPa deduced from resistance measurements identifying the pressure critical point. The apparent difference is likely originated in an underestimation of the applied pressure combined with the small pressure induced differences in an already broad low temperature resistance curve (fig 4.22 in Ref. 9). Our white line peak energy evolution reflects an environment gradually incrementing at 20 K the number of more symmetric octahedra accompanying the transition from insulator to higher conductivity surroundings as the overlap of O 2p and Ni 3d bands increases.

We also follow octahedral shrinking by the analysis of the pseudoradial distribution at 20 K (Fig. 8 (B)). The change of bond lengths up to ~8.1±0.6 GPa where the insulator to metal transition takes place, is in agreement with the precedent discussion. It is the same situation with LaMnO3 [49] where the weak slope of the initial points amounts to a reduction of octahedral tilting and a systematic decrease of the charge disproportionate as a consequence of the compound effect of pressure and temperature driven localization. This trend abruptly changes at the temperature of the insulator-metal transition where the increase in the effective $O_{2P}$ bandwidth (W) closes the charge transfer gap.[22, 23, 49] Increasing pressure signals a net majority of undistorted octahedra in a fashion similar to findings in under pressure $LaMnO_3$ (Ref. 49) leading the system into a multiphase region. In this region, Fig. 8(C), sublattice distortions are found sustaining superexchange angles ranging from orthorhombic to rhombohedral.[26] This last space group starts prevailing at about 27 GPa where there is an enhanced symmetrization. Another break in the slope at about 31 GPa yields the gradual onset of a region with a rhombohedral primary pattern compatible with the R-3c space group.

### *v) High pressure dependence of the Sm $L_3$-edge in $SmNiO_3$ at 300 K and 20 K*

In previous sections ***iii)*** and ***iv)*** we discussed structural changes taken place in the first oxygen coordination shells at the Ni-edge. Fig 6(B) and Fig 8(B) also reveal that in the pseudo radial distribution functions at higher R there are strong changes in shells at distances beyond nearest neighbor oxygen. It involves Rare-Earth ions that either at 300 K or 20 K denote radial distribution oscillations at intermediate pressures becoming damped in the rhombohedral phase. It suggests strong pressure induced distortions for the dodecahedral Sm cage.

This motivate us the study of the $L_3$ Sm edge under pressure. As it was stated in ***ii)***, we established from temperature dependent XAS absorption five different scattering paths accounting for Sm-O bond distances. They are convoluted in an asymmetric first shell envelope. Considering these five paths simultaneously allows, unlike the average weighted single bond model used in Ni-O analysis, choices for correlations of different Sm-O bond lengths that correlate independently from each other. At 20 K, from the possible combinations, meaning sharing Eo, R. and $\sigma^{**2}$ ,the association of scattering paths 1 and 5 (Fig. 3, inset (b)) against the other three yields good fits in the lower pressure range below 9 GPa. This "orthorhombic" model generates steep changes in bond length contractions $\Delta r_1$ and $\Delta r_2$ (crosses and dots in Fig. 9 (B)).

A different combination set involving scattering paths 1 and 4 as the two main begins to prevail at about ~27 GPa where it is found $\Delta r_1$ and $\Delta r_2$ converging toward closer values suggesting a tendency toward smaller distortions. More remarkable, the two contracting parameters merge above 30 GPa into only one reflecting, within the experimental error, the higher symmetric environment imposed by the high-pressure induced rhombohedral phase. At intermediate pressures good R fits result in rather chaotic differences for both independent bond lengths. Seen as uncorrelated random variations of contraction parameters (shadowed section in Fig 9 (B)) is in agreement with the conclusions discussed in section ***iii)*** and ***iv)*** using Ni as absorber, where it was found coexisting a range of different superexchange angles.[26]

## Conclusions

Summarizing, we discussed temperature dependent X-ray absorption measurements of $SmNiO_3$ in the antiferromagnetic, paramagnetic, and metallic phases between 20 and 600 K at the K-Ni and $L_3$-Sm edges. These spectra, in agreement with earlier XAS measurements for other $RNiO_3$ (R=Rare earth), allowed to select main scattering paths, and thus the model, used in the high-pressure analysis. We commented on the pre-edge band origin, and point to post-edge absorption increments as a useful tool for determining the transition to the metallic phase concomitant to the reduction of electron-phonon interactions.

Pressure dependent room temperature Ni white line peak energies have an abrupt valence discontinuity at $T_{MI}$ ~3.10±0.04 GPa from non-equivalent $Ni^{3+\delta}$ + $Ni^{3-\delta}$ charge-disproportionated ions, considered as a net unresolved absorber, turning into $Ni^{3+}$ in the orthorhombically distorted, P*bnm* metal perovskite oxide phase. This finding is consistent with room temperature strong anisotropic changes taking place in the Ni-O bond along with electronic delocalization above $T_{IM}$ earlier found for small Rare-Earth nickelates.[50] At 20 K, we found a much smoother behavior due to localization with the insulator to metal transition shifting up to $T_{IM}$ ~8.1±0.6 GPa.

Applied pressures induce Ni-O-Ni angle increments toward more symmetric $Ni^{3+}$ octahedra in the rhombohedral $R\bar{3}c$ space group which is shared with room temperature metallic $LaNiO_3$. Octahedral bond contraction up to 38 GPa and at 300 K and 20 K show breaks in its monotonic increase at the different structural changes.

The X- ray absorption at Sm $L_3$-edge does not show distinctive features either at 300 K or 20 K up about 35 GPa but the perovskite Sm cage dodecahedral coordinated undergoes critical distortions at intermediate pressures where we found coexistence of distorted octahedra and rhombohedral superexchange angles.

Our findings suggest that the white line pressure dependent anomaly may be used as a sound alternative tool for delineating pressure-temperature phase diagrams. In addition, the new points together with the available data from the current literature, as shown in figure 10, suggest a commonality which is applicable to all P–T phase diagrams in the RNiO3 (R= rare earth≠ La) family of compounds.

.

## Acknowledgements

The authors are grateful to Anna Paula da Silva Soteiro (Laboratorio Nacional de Luz Sincrotron (LNLS), Campinas, SP, Brazil), for her assistant at the X-ray absorption beamline. NEM also thanks LNLS for financial assistant and beamtime allocation under projects N° XAFS116941 and XAFS115947. He is also grateful to the CNRS- Conditions Extrêmes et Matériaux: Haute Température et Irradiation (CEMHTI) laboratory and staff members in Orléans, France, for research and financial support.
Funding through Spain Ministry of Economy and Competitivity (Ministerio de Economia y de la Competividad) under Project MAT2013-41099-R is acknowledged by JAA.

## REFERENCES

1. Demazeau G., Macbeuf A., Pouchard M., Hagemuller P. J., 1971 Solid State Chem. **3**, 582.
2. Medarde M. L., 1997 J. Phys. Cond. Matter **9**, 1679.
3. García-Muñoz J. L., Rodríguez-Carvajal J., Lacorre P., and Torrance J. B., 1992 Phys. Rev. B **46**, 4414.
4. García-Muñoz J. L., Rodríguez-Carvajal J., and Lacorre P., 1994 Phys. Rev. B **50**, 978.
5. Alonso J. A., Martínez-Lope M. J., Casais M. T., Aranda M. A. G., and Fernandez-Diaz M. T., 1999 J. Am. Chem. Soc. **121**, 4754.
6. Alonso J. A., García-Muñoz J. L., Fernández-Díaz M. T., Aranda M. A. G., Martínez-Lope M. J., and Casais M. T., 1999 Phys. Rev. Lett. **82**, 3871.
7. Staub U., Meijer G. I., Fauth F., Allenspach R., Bednorz J. G., Karpinski J., Kazakov S. M., Paolasini L., and d'Acapito F., 2002 Phys. Rev. Lett. **88**, 126402.
8. Höck K.-H., Nickisch H., and Thomas H., 1983 Helv. Phys. Acta **56**, 237.
9. Lengsdorf R. G., 2005 PhD dissertion, Universität zu Köln.
10. Saito T., Azuma M. , Nishibori E., Takata M., Sakata M., Nakayama N., Arima T., Kimura T., Urano C., and Takano M., 2003 Physica B **329**, 866,
11. Acosta-Alejandro M., Mustre de León J., Medarde M., Lacorre Ph., Konder K., and Montano P. A., 2008 Phys. Rev. B 77, 085107.
12. Medarde M., Dallera C., Grioni M., Delley B., Vernay F., Mesot J., Sikora M., Alonso J. A., and Martínez-Lope M. J., 2009 Phys. Rev. B **80** 245105.
13. Medarde M., Lacorre P., Conder K., Rodriguez.Carvajal J., Rosenkranz S., Fauth F., and Furrer A., 1998 Physica B **241-243**, 751.
14. Medarde M., Lacorre P., Conder K., Fauth F., and Furrer A., 1998 Phys. Rev. Lett. **80**, 2397.

15. Torrance J. B., Lacorre P., Nazzal A. I., Ansaldo E. J., and Niedermayer C., 1992 Phys. Rev. B **45** 8209.
16. Mroginski M. A., Massa N. E.,Salva H., Alonso J. A., and Martínez-Lope M. J., 1999 Phys. Rev. B **60**, 5304.
17. Jaramillo R., D. Ha S. D., Silevitch D. M., and Ramanathan S., 2014 Nature Physics **10**, 304.
18. de la Cruz F. P., Piamonteze C., Massa N. E., Salva H., Alonso J. A., Martínez-Lope M. J., and Casais M. T., 2002 Phys. Rev. B 66, 153104.
19. Rodriguez-Carvajal J., Rosenkranz S., Medarde M., Lacorre P., Fernandez-Diaz M. T., Fauth F. ,and Trounov V., 1998 Phys. Rev. B **57**, 456.
20. Lacorre J. B., Lacorre P., Nazzal J., Ansaldo E. J.,and Niedermeyer C, 1992 Phys. Rev. B **45**, 8209.
21. Alonso J. A., Garcia-Muñoz J. L., Fernandez-Diaz M. T., Aranda M. A. G., Martinez-Lope M. J.,and Casais M. T., 1999 Phys. Rev. Lett. **82** 3871.
22. Obradors X., Paulius L. M., Maple M. B., Torrance J. B., Nazzal A. I., Fontcuberta J., Granados X., 1993 Phys. Rev. B **47**, 12353.
23. Canfield P. C., Thompson J. D., Cheong S-W, Rupp L. W., 1993 Phys. Rev. B **47**, 1257.
24. Medarde M., Mesot J., Lacorre P., Rosenkranz S., Fischer P., and Gobrecht K., 1995 Phys. Rev. B **52**, 9248.
25. Lengsdorf R., Barla A., Alonso J. A., Martinez-Lope M. J., Micklitz H., and Abd-Elmeguid M. M., 2004 J. Phys.: Condens. Matter **16,** 3355.
26. Amboage M.,2003 PhD Thesis, Universidad del País Vasco.
27. Lacorre P., Torrance J. B., Pannetier J., Nazzal A. I., Wang P. W., and Huang T. C., 1991 J. Solid State Chem. **91**, 225.
28. Mortimer R., Weller M. T., Henry P. 1999 Physica B **271** 173.
29. Amboage M., Hanfland M., Alonso J. A., and Martínez-Lope M., J. 2005 J. Phys. Cond. Matter **17**, S783.

30. Ramos A. Y., Piamonteze C., Tolentino H. C. N., Sousa-Neto N. M., Bunau O., Joly Y., Grenier S., Itié J.-P, Massa N. E., Alonso J. A., and Martínez-Lope M. J., 2012 Phys. Rev B **85** 045102.
31. Alonso J. A., Martinez-Lope M. J., and Hidalgo M. A., 1995 J. Solid State Chem. **116**, 146.
32. H. C. N. Tolentino, A. Y. Ramos, M. C. M. Alves, R. A. Barrea, E. Tamura, J. C. Cezar, and N. Watanabe, 2001 J. Synchrotron Radiat. **8**, 1040
33. Cezar J. C., Souza-Neto N. M., Piamonteze C., Tamura E., Garcia F., Carvalho E. J., Neueschwander R. T., Ramos A. Y., Tolentino H. C. N., Caneiro A., Massa N. E., J. Martinez-Lope M. J., Alonso J. A., and Itié J.-P, 2010 J. Synchrotron Radiat. **17**, 93.
34. Piamonteze C., Tolentino H. C. N., Ramos A. Y., Massa N. E., Alonso J. A., Martínez-Lope M. J., and Casais M. T., 2005 Phys. Rev. B **71**, 012104.
35. Ragan D. D., Gustavsen R., and Schiferl D., 1992 J. Appl. Phys. **72** 5539.
36. Weinstein B. A., 1986 Rev. Sci. Instrum. **57** 910.
37. Chijioke A. D. , Nellis W. J., Soldatov A., and Silvera I. F., 2005 J. Appl. Phys. **98** 114905.
38. Klotz S., Chervin J-C, Munsch P., and Le Marchand G., 2009 J. Phys. D Appl. Phys. **42** 75413.
39. Ravel B. and Newville M., 2005 J. Synchrotron Rad.**12**, 537.
40. Modrow H., Bucher S., Rehs J. J., and Askudinov A. L., 2003 Phys. Rev. B **67**, 035123.
41. Gougoussis C., Calandra M., Seitzzmen A., Brouder H., Shukla A., Mauri F., 2009 Phys. Rev. B **79**, 045118.

42. Vedrinskii R. V., Kraisman V. L.; Novakovich A. A., Demekhim Ph. V., and Urazhdin S. V., 1998 J. Phys. Cond. Matter **18**, 9561.

43. Piamonteze C., Tolentino H. C. N., and Ramos A. Y., Nucl. Instrum. Methods 2006 Phys. Res., Sect. A **246**, 151.

44. Granados X., Fontcuberta J., Obradors X., Mañosa L., and Torrance J. B., 1993 Phys. Rev. B **48**, 11 666.

45. Medarde M., Fontaine A., Garcia-Muñoz J. L., Rodriguez-Carvajal J., de Santis M., Sacchi M., Rossi G., and Lacorre P., 1992 Phys. Rev. B **46**, 14975.

46. Zhou J-S, Goodenough J. B., and Dabrowski B., 2005 Phys. Rev. Lett. **98**, 127204.

47. Henry P, F., Weller M. T. and Wilson C. C., 2002 Chem. Mater. **14**, 4104.

48. de Groot F., Vanko G. and Glatzel P., 2009 J. Phys. Condens. Matter **21** 104207.

49. Ramos A. Y., Tolentino H.C.N., Souza-Neto N. M., ITié J-P, Morales L. and Caneiro A., 2007 Phys. Rev. B **78**, 052103

50. Alonso J. A., Martínez-Lope M. J., Casais M. T., Garcia Muñoz J. L., Fernández-Díaz M. T., and Aranda M. A. G., 2001 Phys. Rev. B **64**, 094102.

51. N. F. Mott, 1949 Proc. Phys. Soc. (London) **A62**, 416.

52. J. Ruppen, J. Tysler, O. E. Pell, S. Catalano, M- Gilbert, J. Mravlje, J.-M Tricone, A. Georgees and D. van der Marel. 2015 Phys. Rev. B **92,** 155145

## 53.**FIGURE CAPTIONS**

**Figure 1.** (color online) Normalized absorption spectra at the Nickel K-edge as a function of temperature at ambient pressure. For better view the spectra have been displaced vertically. Features named after, *A*: pre-edge; *B*: while line, and *C*: post-edge shoulder, are discussed in the text. Inset: temperature dependence of Ni-O bond distances from pseudo distribution function fits

**Figure 2.** (color online) **(A)** Normalized Nickel K pre-edge XANES at 20 K; squares: data; full lines: fit. Inset (a): pre-edge bands at 20 K antiferromagnetic, paramagnetic 300 K, and metallic 600 K phase respectively. Inset (b): $A_1$ peak position from first derivative suggesting faint softening in the metallic phase. **(B)** Nickel K post-edge spectra from 20 K to 600 K. Inset: enhanced spectra showing the compound response at 20 K ((antiferromagnetic), above ~230 K (paramagnetic insulator), and above $T_{MI}$ ~400 K in the metal oxide phase up to 600 K.

**Figure 3.** (color online) Normalized absorption spectra at the Sm $L_3$-edge as a function of temperature at ambient pressure. For better viewing, the spectra have been displaced vertically. Inset (a): XANES detail at 20 K (antiferromagnetic insulator), 300 K (paramagnetic insulator), and 600 K (metal oxide). Inset (b): Calculated bond distances from Sm-O scattering path fits

**Figure 4.(A)** (color online) Normalized absorption spectra at the Nickel K-edge as a function of pressure at 300 K (dashed lines are guides for the

eye)**.(B)**:Nickel K post-edge spectra up to 38 GPa and at 300 K. Inset: Enhanced spectra showing the compound passage into the metallic phase.

**Figure 5.** (color online) Pressure dependent of normalized Nickel K pre-edge XANES at selected pressures. Left column: 300 K; right column: 20 K.

**Figure 6**(color online)**.(A)** White line Ni-K edge pressure dependent peak energy at 300 K **(B)** $k^2$ weighed absolute value of pressure dependent Fourier transform EXAFS oscillations at 300 K, **(C)** Pressure dependence of the average Ni-O bond length contraction from the EXAFS fits at 300 K; full line is guide for the eye.

**Figure 7.** (color online) **(A)** Normalized absorption spectra at the Nickel K-edge as a function of pressure at 20 K (dashed lines are guides for the eye) **(B)** Nickel K post-edge spectra up to 38 GPa and at 20 K. Inset: Enhanced spectra showing the compound passage into the metallic phase.

**Figure 8.** (color online) **(A)** White line Ni-K edge pressure dependent peak energy at 20 K, **(B)** $k^2$ weighed absolute value of pressure dependent Fourier transform EXAFS oscillations at 20 K, **(C)** Pressure dependence of the average Ni-O bond length contraction from the EXAFS fits at 20 K; full line is guide for the eye.

**Figure 9.** (color online) **(A)** Normalized absorption spectra at the Sm $L_3$-edge as a function of pressure at 20 K**.** For better viewing spectra have been displaced vertically. **.(B)** $k^2$ weighed absolute value of pressure dependent Fourier transform EXAFS oscillations at 20 K, **(C)** Pressure dependence of $\Delta r_1$ and $\Delta r_2$ (crosses and dots) average Sm-O bond contractions from the EXAFS fits at 20 K.

**Figure 10.** (color online) P-T phase diagram for $PrNiO_3$ (empty diamonds: Ref, 23; empty triangles: Ref. 22), $NdNiO_3$ (full diamonds: Ref. 23; full triangles: Ref. 22 and $SmNiO_3$ (dots: Ref. 9, triangle: Ref. 28, squares: this work). Full line is a cubic polynomial fit intended as a guide for the eye.

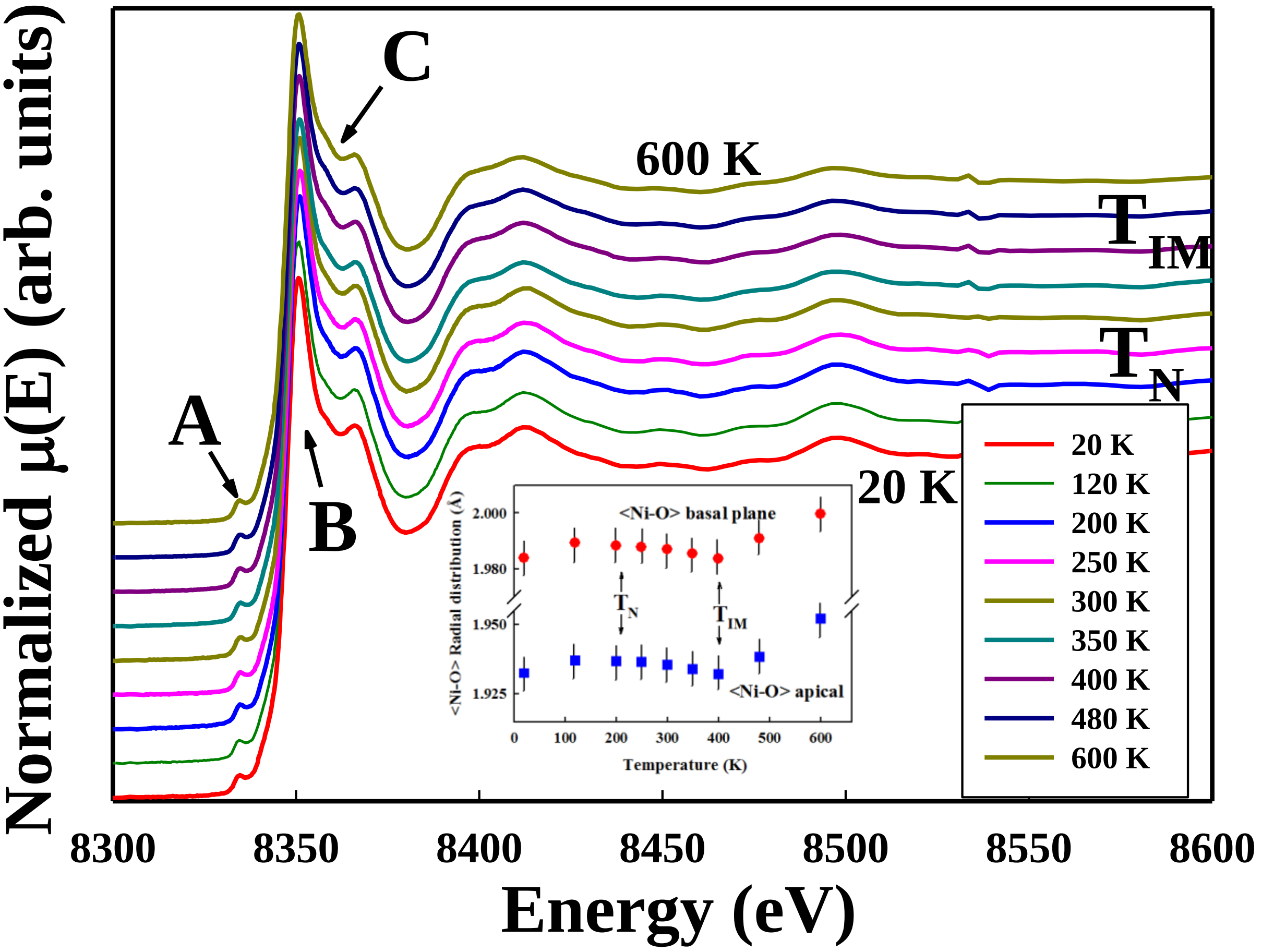

**Figure 1**
**Massa et al**

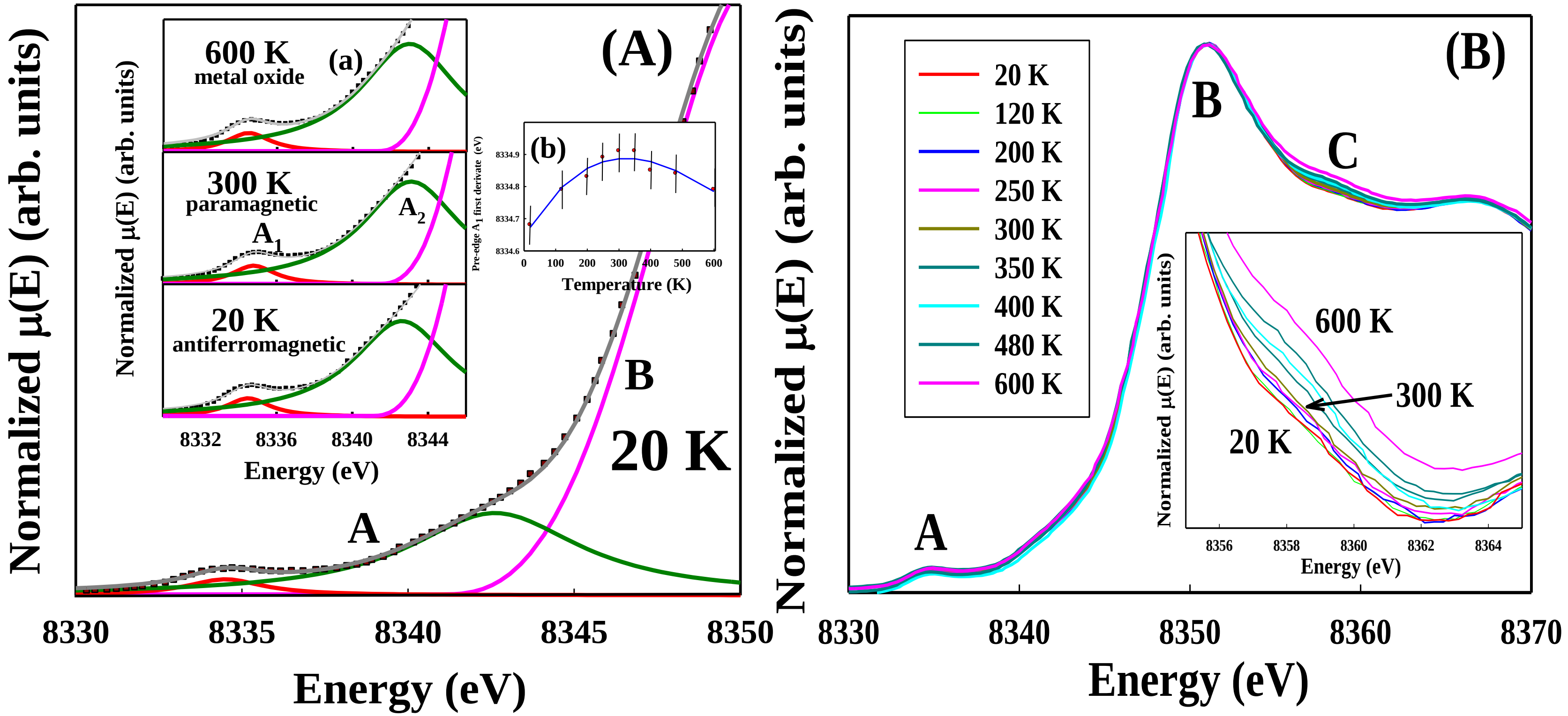

**Figure 2**
**Massa et al**

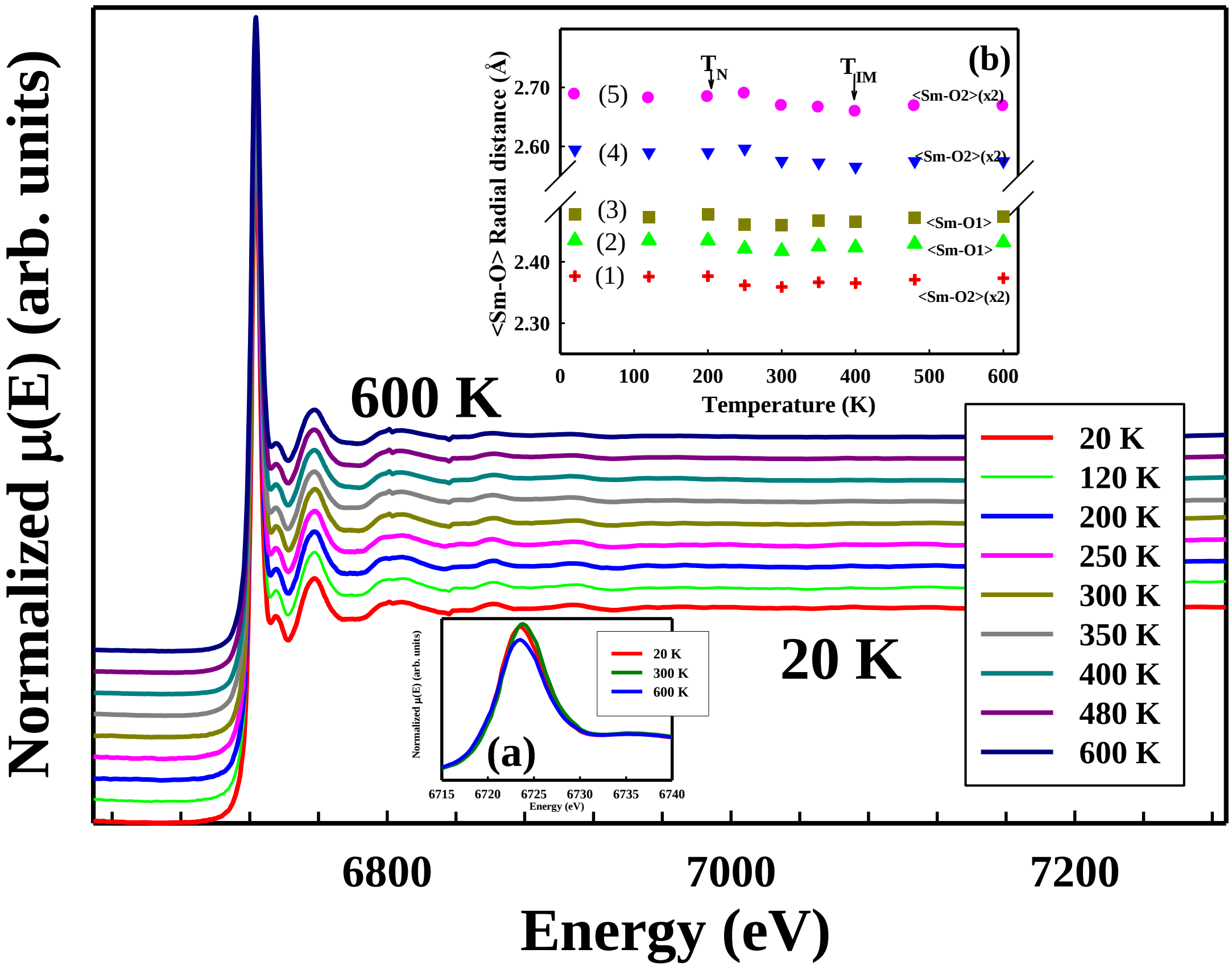

# Figure 3
# Massa et al

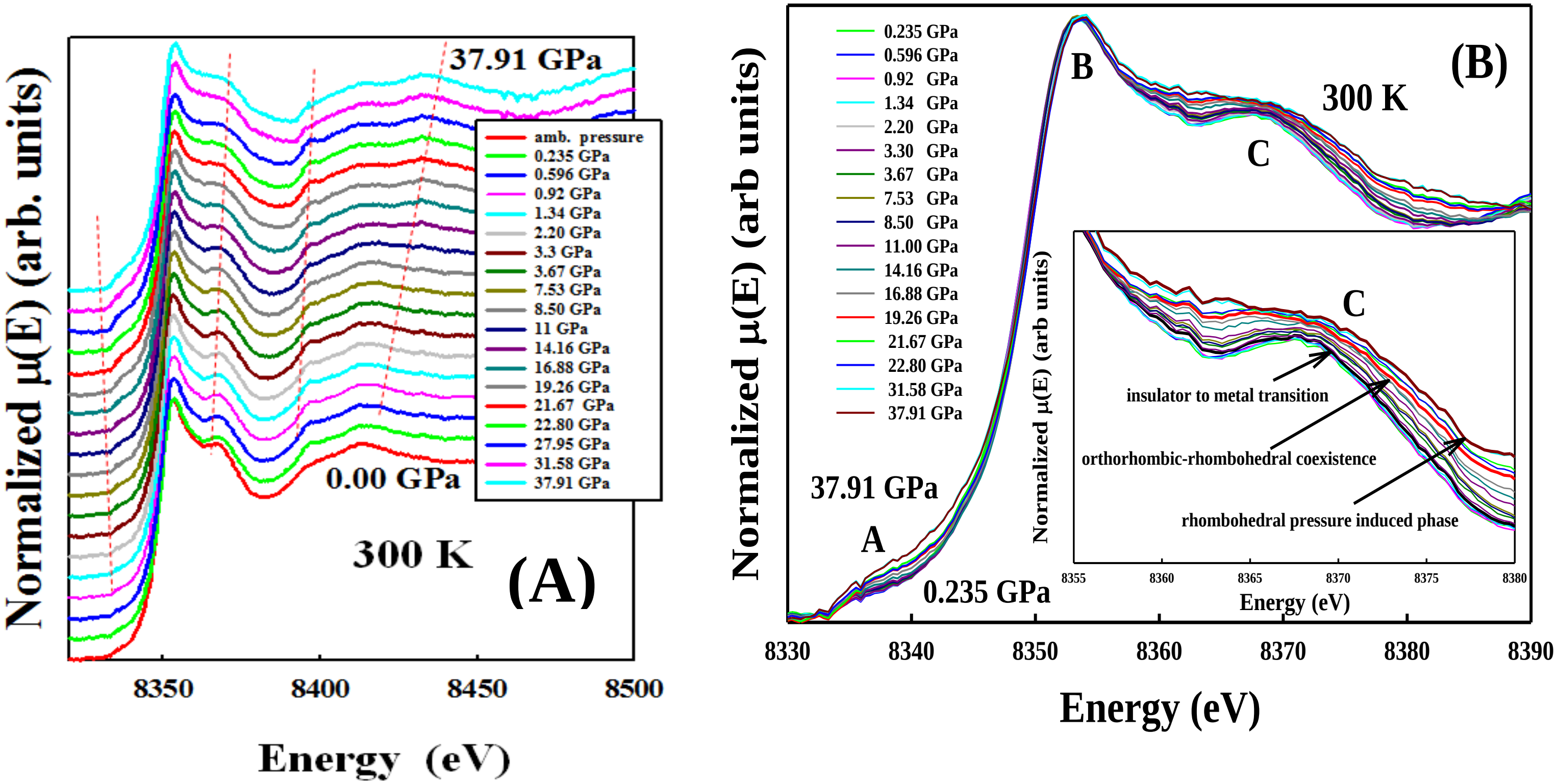

# Figure 4
# Massa et al

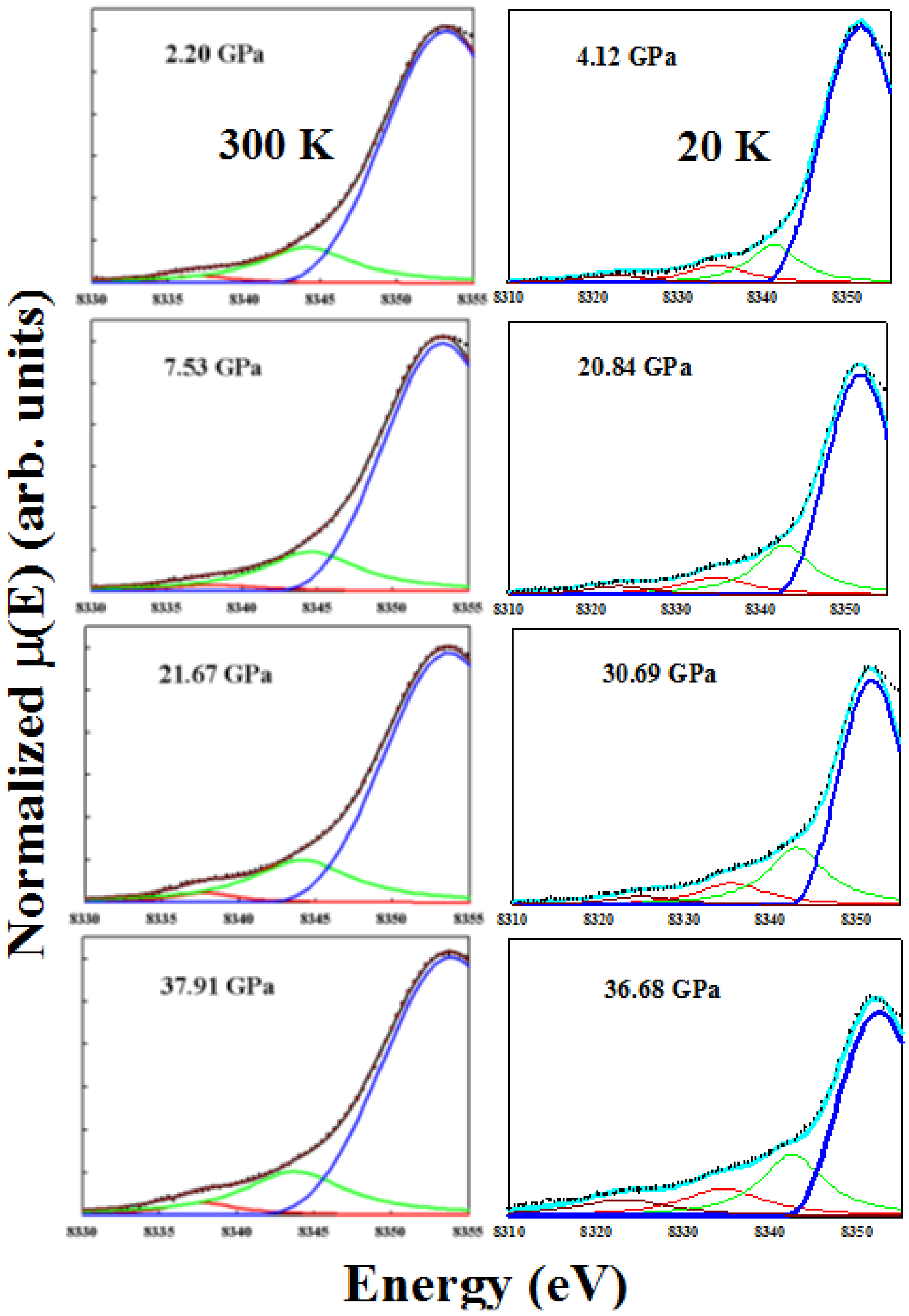

# Figure 5
# Massa et al

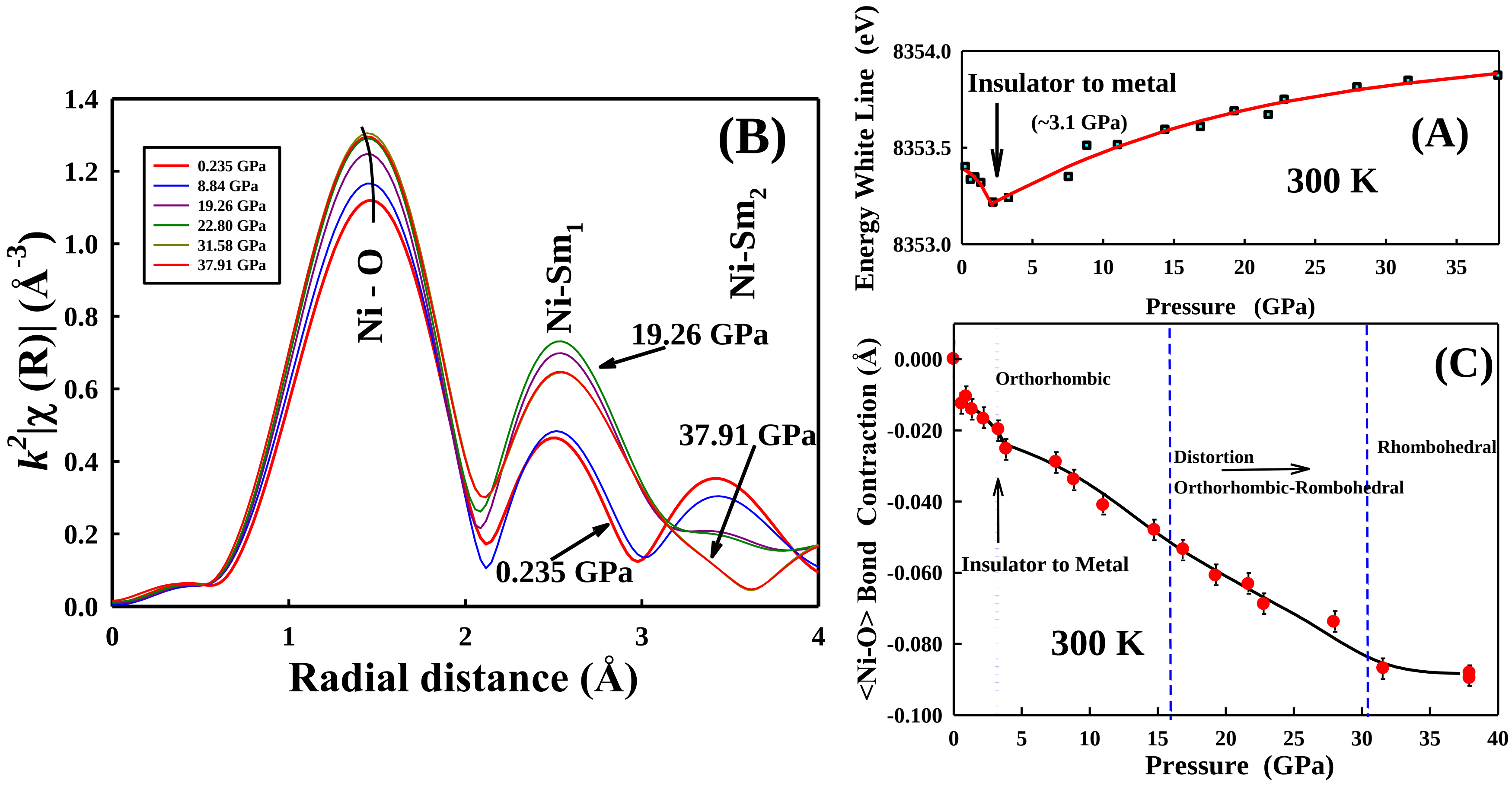

**Figure 6**
**Massa et al**

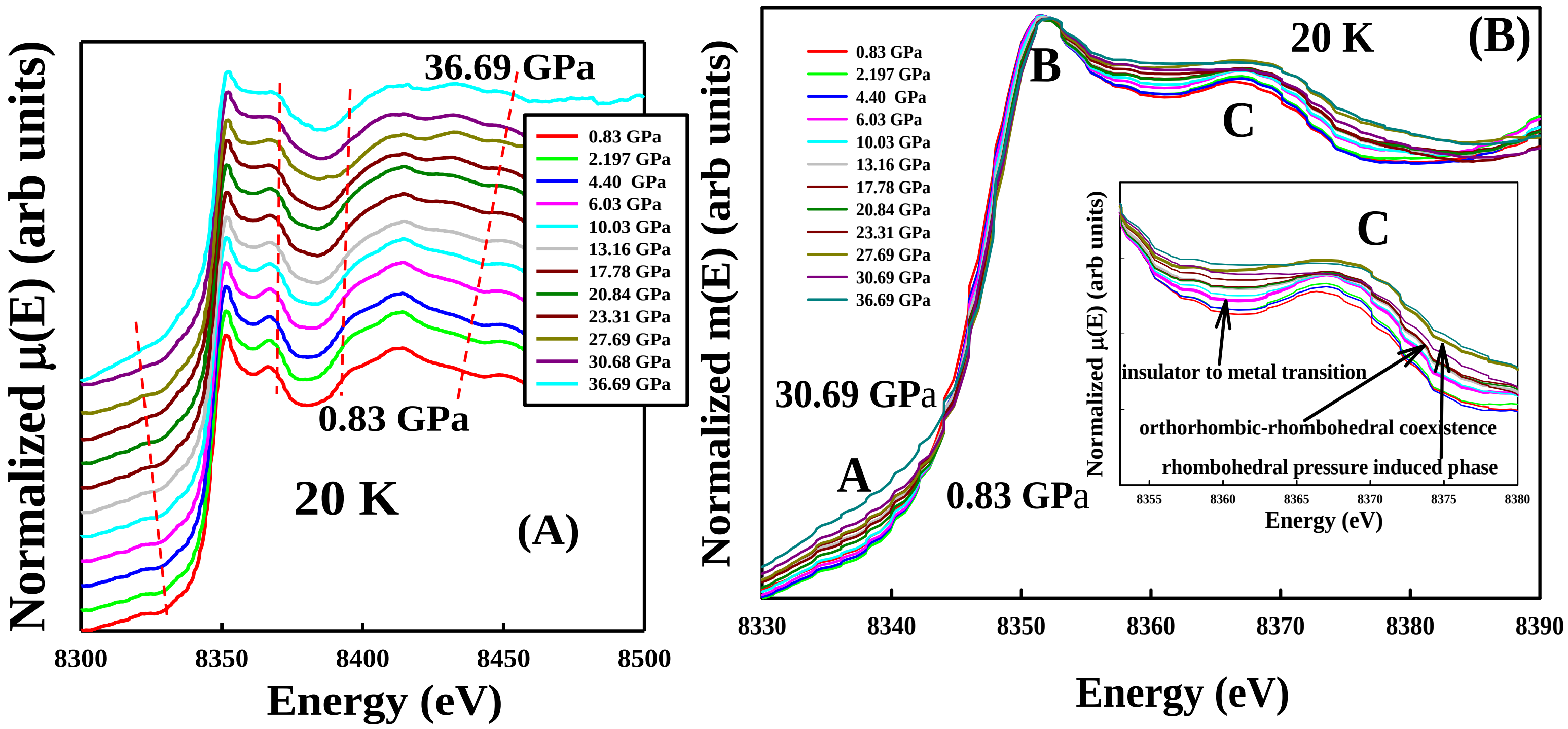

# Figure 7
# Massa et al

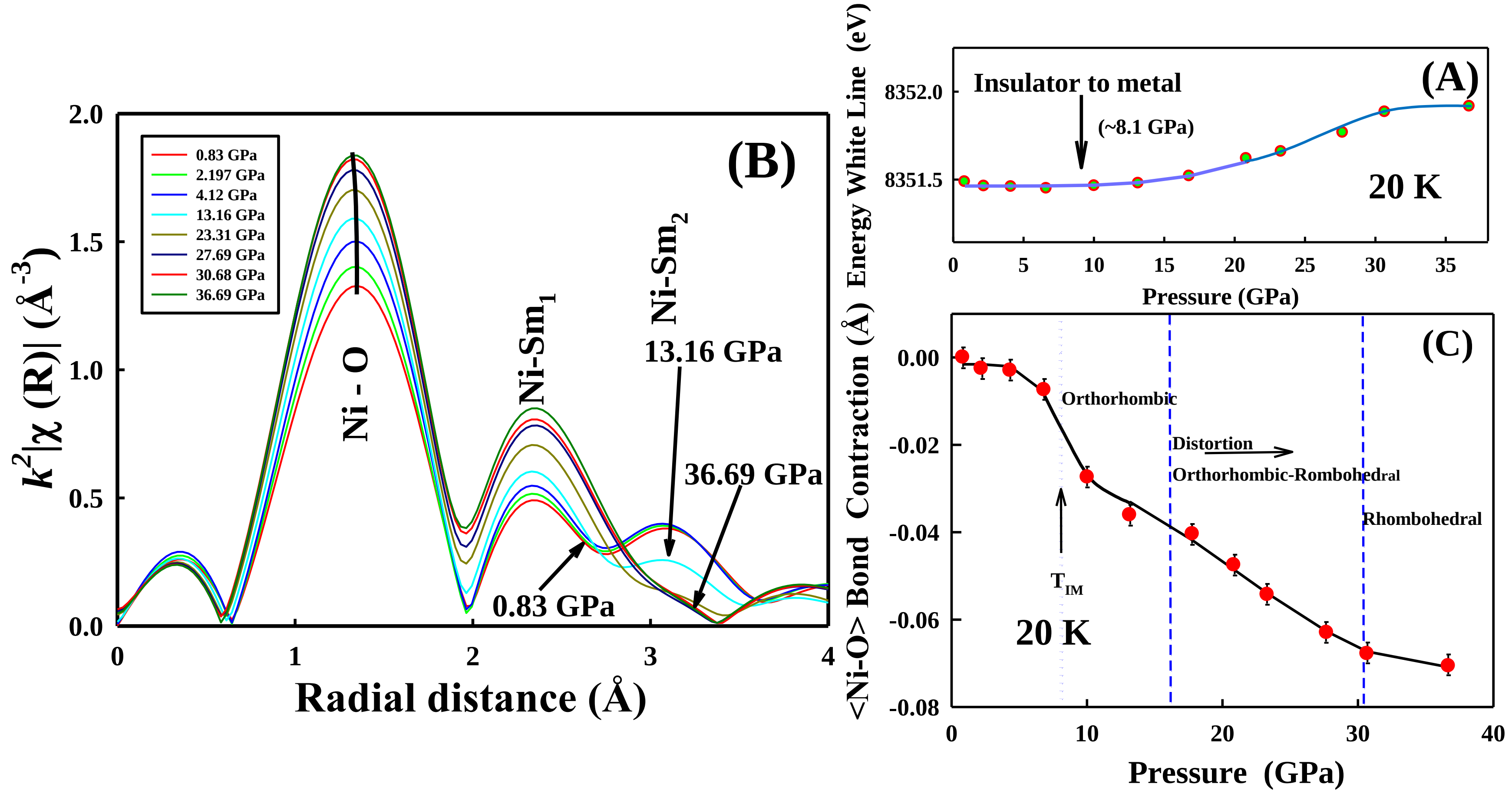

# Figure 8
# Massa et al

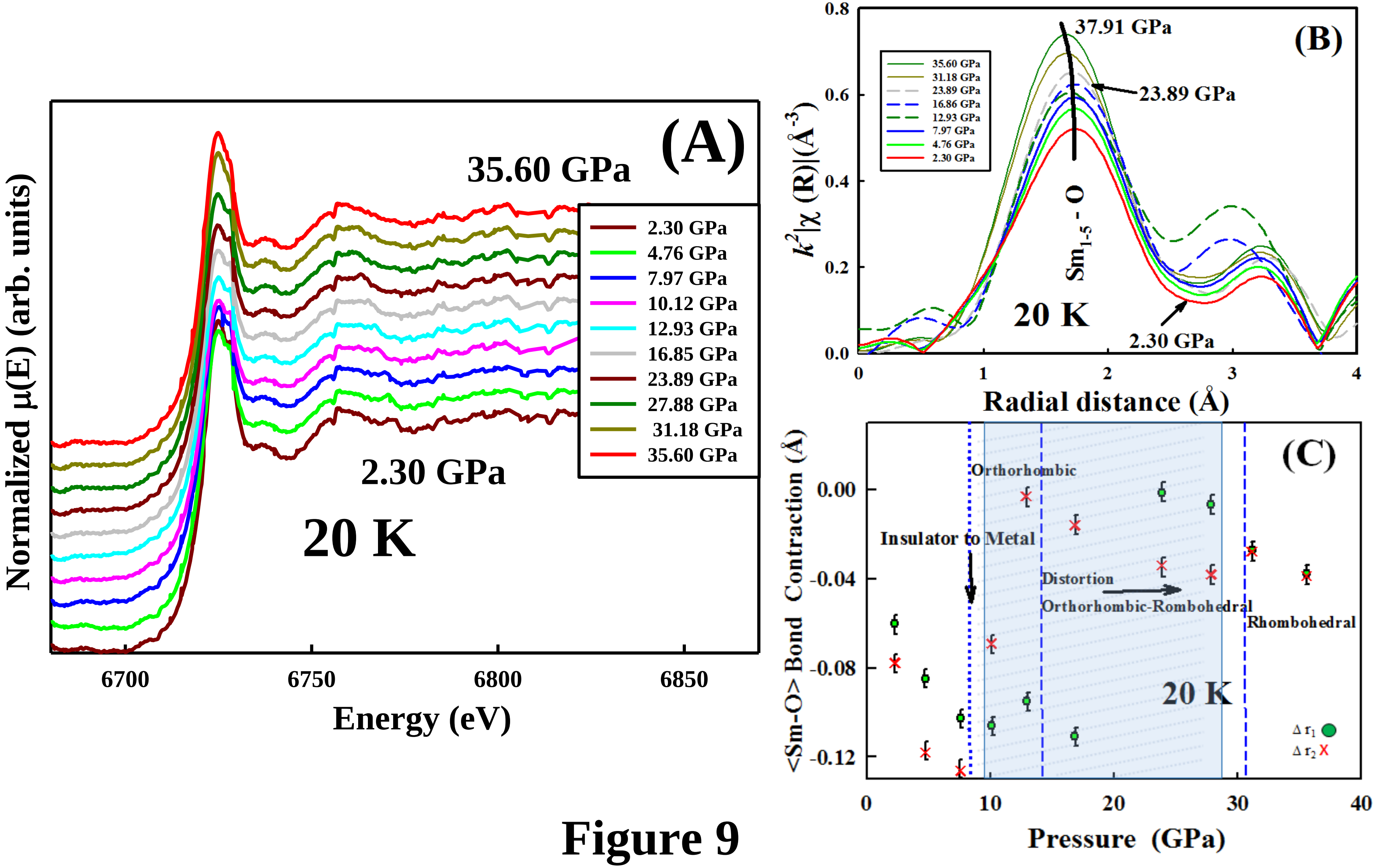

# Figure 9
# Massa et al

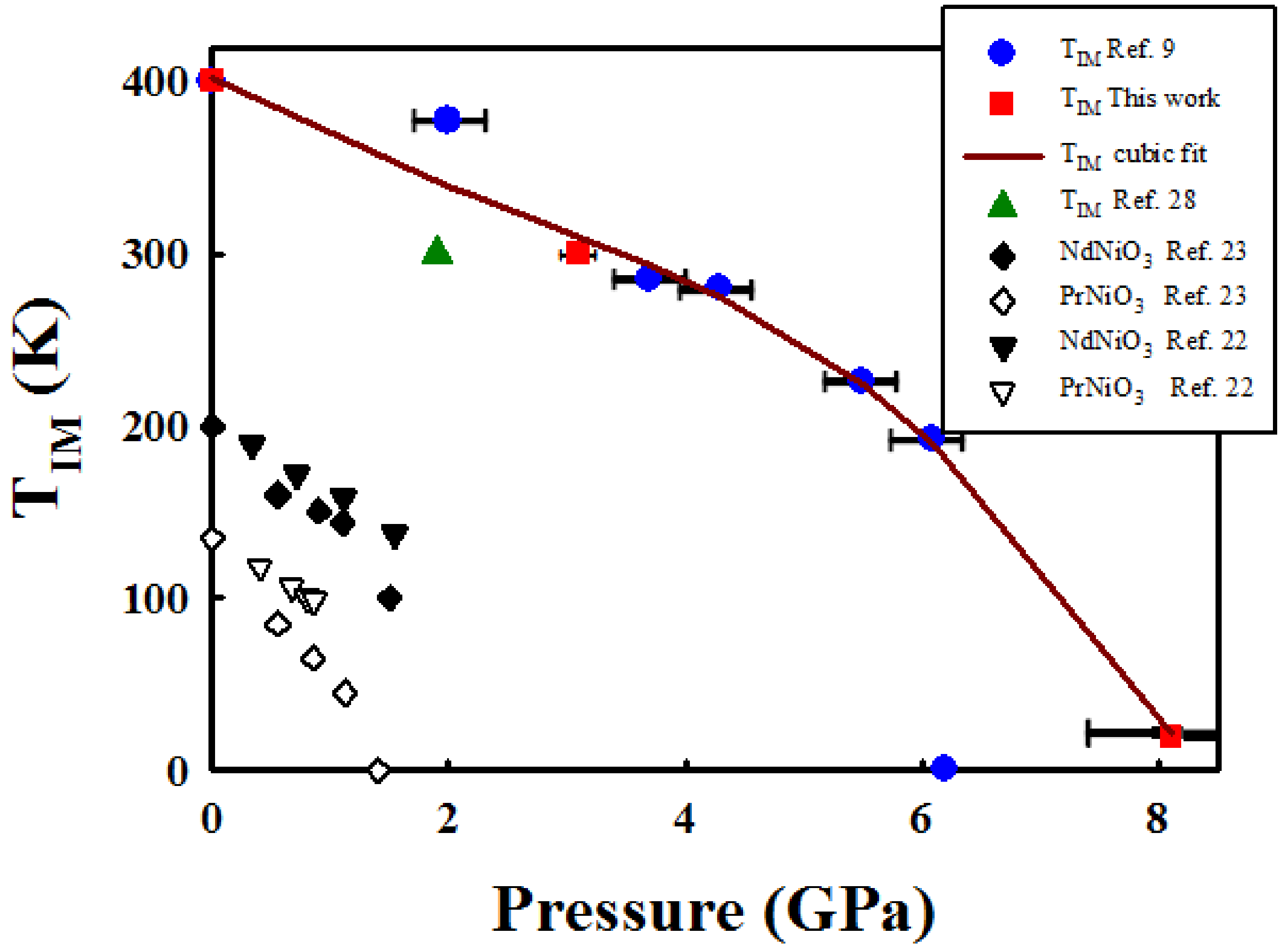

**Figure 10**
**Massa et al**